# Shoreline Responses to Rapid Water Level Increases in Lake Michigan

**Hazem U. Abdelhady[1,2], Cary D. Troy[1], Longhuan Zhu[3], Pengfei Xue[3,4,5], Guy Meadows[3] and Chin H. Wu[6]**

[1]Purdue University, Lyles School of Civil and Construction Engineering, West Lafayette, IN, USA, [2]Cooperative Institute for Great Lakes Research, University of Michigan, Ann Arbor, Michigan, USA, [3]Great Lakes Research Center, Michigan Technological University, Houghton, MI, USA[4]Department of Civil, Environmental and Geospatial Engineering, Michigan Technological University, Houghton, MI, USA, [5]Environmental Science Division, Argonne National Laboratory, Lemont, IL, USA, [6]Department of Civil and Environmental Engineering, University of Wisconsin-Madison, Madison, WI, USA

Corresponding author: Hazem Abdelhady ([huhady@umich.edu](mailto:huhady@umich.edu))

*440 Church St, Ann Arbor, MI 48109

## Abstract

High-resolution multispectral satellite imagery was utilized to quantify shoreline recession at eleven beaches around Lake Michigan during a record-setting water level increase between 2013 and 2020. Shoreline changes during this period ranged from 20 m to 62 m, corresponding to 52-95% of the initial beach widths. Average estimated shoreline erosion across all beaches varied from 1% to 75% of the observed changes, with the remainder attributed to inundation. Significant correlations were found between shoreline erosion and wave-related factors, including offshore wave power, offshore bathymetric slope, storm energy, and potential alongshore sediment transport divergence. In contrast, parameters related to cross-shore transport, such as dimensionless fall velocity, exhibited weak correlations. Additionally, the results underscore the importance of distinguishing between immediately reversible changes (inundation) and morphological changes that could be reversible over longer timescales, when assessing the impact of rising water levels., The findings also suggest that in addition to waves playing a key role in regulating shoreline changes, alongshore sediment transport processes may play a more crucial role in beach erosion during significant water level increases than cross-shore processes, challenging traditional models of beach adjustment to rising waters.

## 1 Introduction

The shoreline position of lakes and oceans is one of key variables for coastal engineers and scientists studying coastal evolution (Douglas & Crowell, 2000; Vos et al., 2019). Shoreline changes can affect natural habitats, biodiversity and ecological balance (Elias & Meyer, 2003; Porst et al., 2019), resulting in significant economic and social impacts (Lent, 2004). Therefore, better understanding of shoreline changes can provide insight to protect human lives and properties, preserve intricate coastal ecosystems, and develop sustainable coastal management (Dean & Dalrymple, 2001b; Lent, 2004).

Water level fluctuations play an important role in shoreline dynamics in coastal areas (Dean & Houston, 2016; Miller & Dean, 2004). Unlike oceans, the Laurentian Great Lakes are largely tideless but still exhibit significant interannual and long-term water level fluctuations. These fluctuations are critical in shaping the shoreline evolution of the Great Lakes (Abdelhady & Troy, 2023b, 2023a; Theuerkauf et al., 2019; Troy et al., 2021). Historically, Great Lakes water levels have followed an 8-10 year oscillation pattern, though hydrological forecasts suggest that climate change may increase the frequency of these cycles (Gronewold & Rood, 2019; Hanrahan et al., 2009; Troy et al., 2021). These interannual water level changes are crucial in shaping the long-term evolution of Lake Michigan's natural and engineered beaches (Abdelhady & Troy, 2023a, 2023b; Mattheus et al., 2022). Recently, Lake Michigan experienced a dramatic water level increase of over 2 meters from 2013 to 2020 (~ 300 mm/year), compared to the global ocean's rise of 3.4 mm/year (NOAA, 2021). These rapid changes have caused widespread coastal impacts, including shoreline erosion, habitat loss, dune and bluff recession, infrastructure damage, and coastal flooding (Jose et al., 2022; Theuerkauf et al., 2019; Theuerkauf & Braun, 2021; Troy et al., 2021). Nevertheless, the spatial variability of these shoreline changes is not yet well understood.

Shoreline changes are commonly associated with factors related to hydrodynamics, morphology, and/or sediment transport processes. Wave energy has been linked to coastal and shoreline changes, with increased wave energy resulting in more erosion (Abdelhady & Troy, 2023a; Davidson et al., 2013; Vitousek & Barnard, 2015; Yates et al., 2009). Storm wave energy also explains shoreline change variability during periods with similar wave energy (e.g. Karunarathna et al. (2014)). Wave direction, specifically bi-directional wave climates, plays a critical role to increase erosion risks and the variability of shoreline changes (Wiggins et al., 2019). In regions with low temperatures, such as the Great Lakes, ice cover can have great impacts on the shoreline. More days without ice cover can led to more wave energy reaching the beach, thereby increasing erosion (BaMasoud & Byrne, 2012).

Net longshore sediment transport affects the distribution of sediments along the coast (Vitousek & Barnard, 2015). Positive (negative) values indicate sediment surplus (deficit) in the system, which indicates shoreline accretion (erosion) with values close to zero indicating no net change. Morphological characteristics such as beach slope can significantly influence shoreline change. Mild slopes dissipate significant wave energy through bottom friction and wave breaking, while steeper slopes allow for waves to break closer to shore (Putnam & Johson, 1949).

Non-dimensional factors like dimensionless fall velocity ($\Omega$), which is defined as the wave height divided by wave period and sediment fall velocity, have been linked to beach typology, erosion, and shoreline changes. Low values of dimensionless fall velocity are more likely to be associated with onshore sediment transport, resulting in beach accretion, and vice versa (Dean & Dalrymple, 2001b). Kraus et al. (1991) found that wave steepness divided by dimensionless fall velocity to the third power ($S_o/\Omega^3$) is inversely proportional to beach erosion. Last but not least, the Bruun Rule slope is often applied to estimate oceanic shoreline changes due to long-term sea level rise (Dean & Houston, 2016). It is implicitly influenced by wave height and bathymetry through the depth of closure (Nicholls et al., 1998). While all these factors have been applied to explain oceanic coastal shoreline responses, no comprehensive studies have assessed these factors and their relative influence on shoreline changes, particularly in the context of rapid water level rise, such as those experienced by the Great Lakes of North America.

The objectives of this paper are twofold: (1) to quantify Lake Michigan shoreline changes at select beaches in response to the recent 2013-2020 water level increase; (2) to investigate linkages between these observed shoreline changes and explanatory or correlative factors related to wave climate, morphology, and sediment transport. By addressing these objectives, the aim is to identify the most vulnerable areas and to provide a better understanding of the processes driving shoreline changes in response to water level increases.

## 2 Methods

### 2.1 Study beaches

Eleven beaches were selected along the Lake Michigan shoreline to represent a diverse range of geographic locations and physical characteristics. The selected sites encompass a mix of shoreline conditions with varying degrees of wave exposure and fetch. Beaches with significant human modifications, such as extensive beach nourishment or shoreline hardening, were avoided to focus on areas with more natural shoreline dynamics. Additionally, only beaches with clear, unobstructed satellite imagery during the two target periods (June–July 2013 and June–July 2020) were included, avoiding areas where trees or frequent cloud cover might obscure the shoreline (Table 1 and Fig. 1).

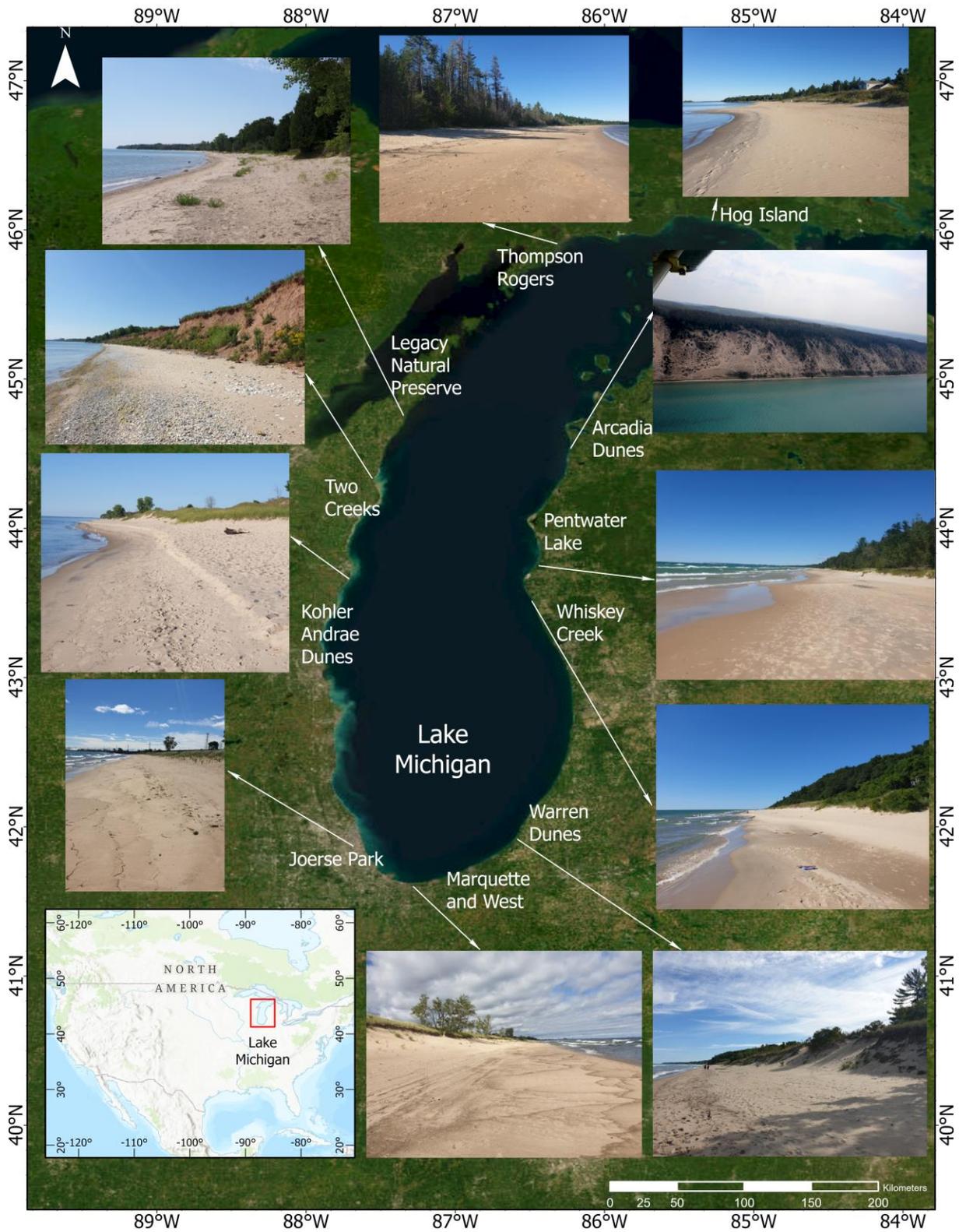

**Figure 1**. The eleven selected natural beaches along the Lake Michigan shoreline are shown. The images were taken during a site visit to all the sites in August 2022, except for the Arcadia Dunes aerial image, which was downloaded from (https://portal1-

**Table 1**. General characteristics of the chosen study sites. Dune heights and beach width were calculated based on the 2012 topobathymetric LiDAR. The beach width was calculated as the distance from the 2012 shoreline (wet-dry line) to the dune toe.

| | Latitude and longitude | Sediment type | Beach type | Sediment size (D50, mm) and coefficient of uniformity | Backshore conditions | 95th Percentile significant wave height (m) | Alongshore length (m) | Beach width 2012/ 2013 (m) | Lake Michigan shore |
|---|---|---|---|---|---|---|---|---|---|
| Whiskey Creek (WC) | 43°31'58.36"N, 86°29'24.93"W | medium-grained sand | exposed | 0.35, 1.56 | vegetated sandy dunes, 15 m in height | 2.04 | 2312 | 81 | east |
| Pentwater Lake (PL) | 43°45'26.64"N, 86°27'15.80"W | medium-grained sand | exposed | 0.28, 1.44 | vegetated sandy dunes, 15 m in height | 1.87 | 2457 | 87 | |
| Arcadia Dunes (AD) | 44°31'52.65"N, 86°13'52.86"W | medium-grained sand | exposed | 0.35, 1.53 | vegetated sandy dunes, 80 m in height | 1.84 | 3128 | 40 | |
| Kohler Andrae Dunes (KA) | 43°39'45.08"N, 87°43'6.04"W | medium-grained sand | exposed | 0.33, 1.77 | vegetated sandy dunes, 5 m in height | 1.74 | 2445 | 60 | |
| Two Creeks (TC) | 44°19'53.75"N, 87°32'29.27"W | coarse-grained sand | semi-embayed | 0.70, 2.82 | vegetated clayey dunes, 11 m in height | 1.53 | 2072 | 25 | west |
| Legacy Nature Preserve (LNP) | 44°45'4.15"N, 87°20'3.89"W | medium-grained sand | semi-embayed | 0.33, 1.83 | no dunes; scattered houses with scattered concrete blocks | 1.74 | 1863 | 33 | |

| Thompson Rogers (TR) | 45°55'5.46"N, 86°18'41.56"W | fine-grained sand | embayed | 0.23, 1.85 | no dunes, scattered trees | 1.78 | 3005 | 66 | north |
| Hog Island (HI) | 46° 4'4.22"N, 85°16'21.54"W | medium-grained sand | semi-embayed | 0.34, 1.49 | vegetated sandy dunes, 4 m in height | 1.13 | 782 | 59 | |
| Warren Dunes (WD) | 41°55'56.08"N, 86°35'25.61"W | medium-grained sand | exposed | 0.34, 1.41 | vegetated sandy dunes, 8-14 m in height | 1.66 | 3122 | 78 | south |
| Marquette and West (MW) | 41°37'19.61"N, 87°14'29.72"W | medium-gained sand | exposed | 0.33, 1.62 | vegetated sandy dunes, 7 m in height | 1.35 | 5080 | 53 | |
| Jeorse Park (JP) | 41°38'41.30"N, 87°25'40.74"W | medium-grained sand | embayed | 0.26, 1.81 | vegetated sandy dunes, 4 m in height; some loosen concrete blocks in the middle | 1.30 | 1585 | 44 | |

## 2.2 Shoreline change quantification

Shoreline positions and changes for the selected beaches were quantified using multispectral satellite imagery, following the methods outlined in Abdelhady et al. (2022). Shoreline position was defined as the transition from water to dry land, which is readily distinguishable from multispectral satellite imagery. Planetscope and Rapideye satellite imagery (Planet Labs, 2020) was acquired for the periods from June/July 2013 and June/July 2020, spanning the full period of lake-level rise. RapidEye images are available from 2009 to 2020 with a 5 m pixel size and a 5.5-day revisit time, while Planetscope is available from 2016 to the present with a 3 m pixel size and a 1-day revisit time. The shoreline position and shoreline changes between images were extracted using the shoreline model that was developed by Abdelhady et al. (2022). The model uses the direct difference water index (DDWI) with a simple technique to determine the threshold between the water and land from the DDWI probability density function distribution. The transect-free method developed by Abdelhady et al. (2022) was used to determine the shoreline changes between different images. The average of all these changes was used to reduce the overall uncertainty. The uncertainty in each calculated shoreline change is the sum of half the pixel size of each of the two images used, which in this case is 4 m (1.5 m+2.5 m). The overall uncertainty was calculated as $\frac{4}{\sqrt{n}}$, where n is the number of coupled images used to calculate the average shoreline changes for a given site. Further details about the detection framework, its accuracy, and validation can be found in (Abdelhady et al., 2022).

It is crucial to distinguish between shoreline erosion and inundation, as they represent different processes with distinct implications (Abdelhady & Troy, 2023b, 2023a; Hands, 1984; Vitousek et al., 2017). Inundation refers to shoreline movement due to rising water levels, which can be swiftly reversed when levels recede, a common occurrence in the Great Lakes. In contrast, shoreline erosion involves sediment transport, making it more challenging to reverse. To isolate shoreline erosion from inundation effects, passive flooding—shoreline movement caused by water level changes without morphological alterations—was subtracted from total detected shoreline changes (Figure 2; (Meadows et al., 1997; Vitousek et al., 2017) ). Passive flooding was estimated by multiplying water level changes with beach face slopes, which were derived from cross-shore profiles spaced 20 m apart using USACE 2012 topo-bathy LiDAR data (OCM Partners, 2012). For each profile, the slope was calculated as $\frac{WLE_{2020} - WLE_{2013}}{L_h}$, where $WLE_{2020}$ and $WLE_{2013}$ are the average water level in June 2020 and June 2013, respectively, and $L_h$ is the horizontal distance between these elevations. The mean slope of all profiles along a beach was used to represent that beach's slope.

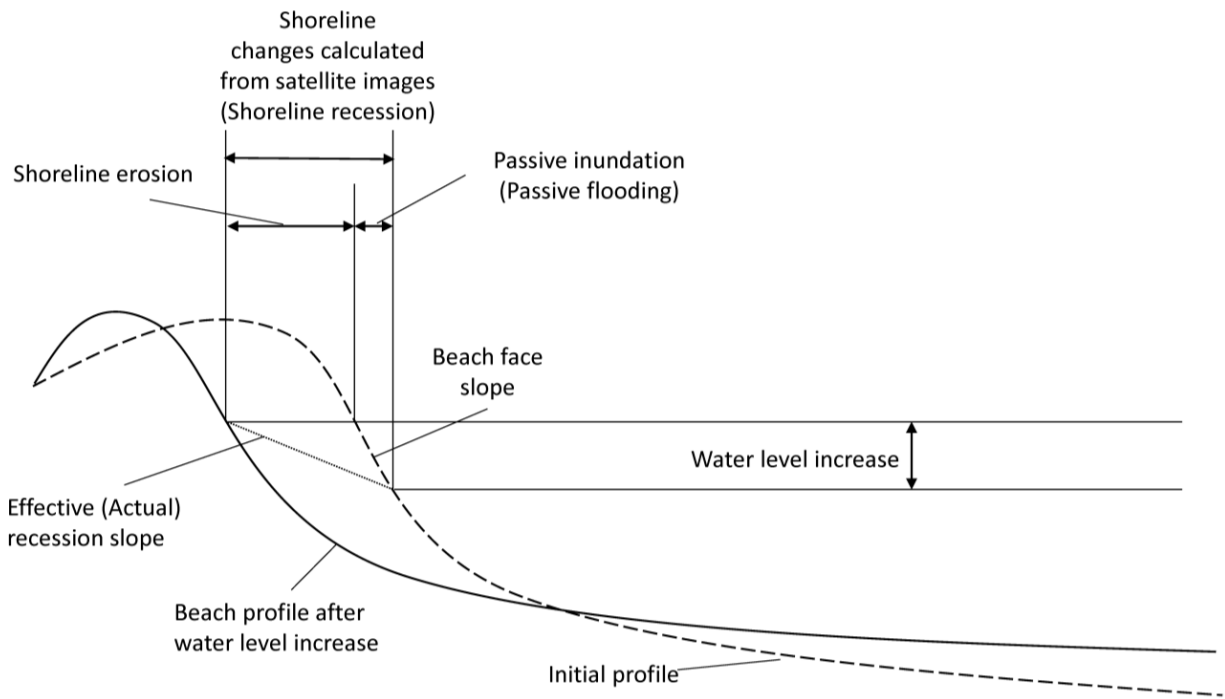

**Figure 2.** Schematic showing the different types of shoreline changes used in this study. Adapted from (Meadows et al., 1997).

### 2.3 Waves

To determine wave climates for the sites, a hindcast model was carried out for Lake Michigan using the physics-based nearshore wave model SWAN v41.01 sssssss(https://swanmodel.sourceforge.io/) (Huang et al., 2021). The grid resolution for the

SWAN model is approximately 1 km. The model was driven by hourly surface wind forcing from the from Climate Forecast System Reanalysis (CFSR, (Saha et al., 2010)) for 1991–2010 and Climate Forecast System Version 2 (CFSv2, (Saha et al., 2014)) for 2011–2020, which have a spatial grid resolution of ~0.3° and ~0.2°, respectively.

To include the effect of ice on the wave field, historical ice fields were incorporated into wave simulations by employing a technique that treated ice-covered areas as lands when ice coverage was higher than 30%. Historical ice data for the lake are accessible via NOAA Great Lakes Ice Atlas and Great Lakes Environmental Research Laboratory (https://www.glerl.noaa.gov/data/ice/). The ice data has a horizontal resolution of 2.5 km for years prior to 2007 and a resolution of 1.8 km for years after 2007, with daily temporal resolution. More details regarding the ice datasets are available (R. Assel et al., 2003; R. A. Assel et al., 2013; Yang et al., 2020).

All the wave grid points were chosen at a distance of approximately 5 km perpendicular to the middle of each beach shoreline, which was a sufficient distance such that most waves were in deep water and were not influenced by nearshore bathymetry. Wave height time series for all the stations from 2013 to 2020 and the wave roses are shown in Figures 3 and 4.

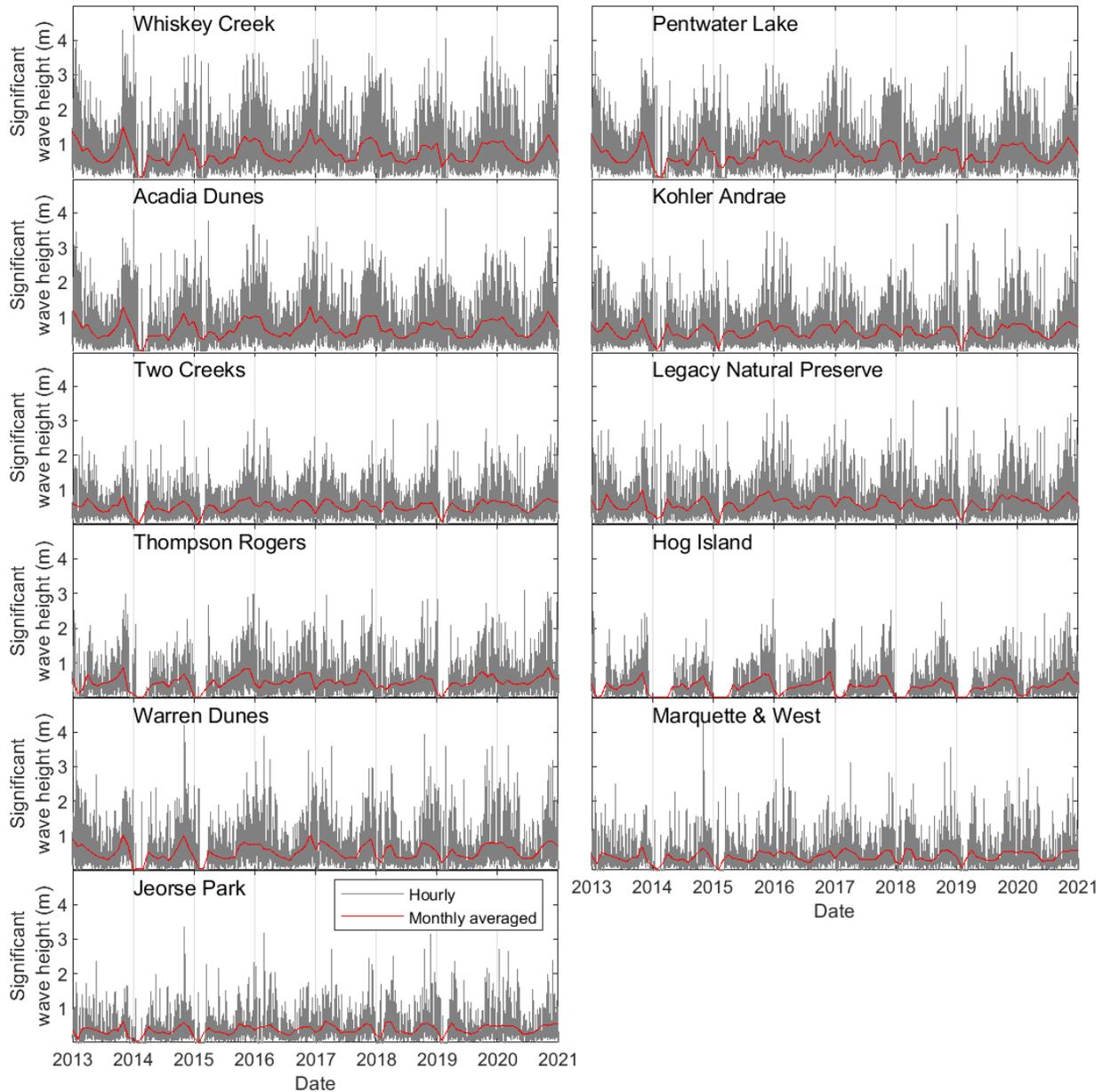

**Figure 3.** Simulated significant wave height time series for all the study locations. Gaps in the wave time series are caused by periods of ice cover.

The wave model was comprehensively validated by comparing the model results with the measured significant wave height and peak wave period at 18 buoy stations in Lake Michigan (Huang et al., 2021). Validation was performed using data from buoy 45002 in the northern basin, available since 1979, and buoy 45007 in the southern basin, available since 1981. Additionally, 10 years of data from 16 additional buoys located in shallower waters along the coast, all within less than 7 km of the shoreline, were used. Four statistical error measures (bias, root mean square error, scatter index, and $R^2$) suggest good performance of the model ($R^2$ ranged from 0.73 to 0.91 for most buoys). Full details about the model validation are referred available in Huang et al. (2021).

### 2.4 Sediment characteristics

Three to six sand samples were collected from the berm and the beach face at each site to determine general sedimentary characteristics (Table 1). The samples were taken 200 m apart in the longshore direction within the extent of the sites. Following the ASTM and EM 1110-2-1906 standards (MacIver & Hale, 1970), dry sieve analysis was performed on all samples to determine the $D_{10}$, $D_{50}$, and $D_{60}$. $D_{10}$ $D_{50}$, and $D_{60}$ are the particle sizes with 10, 50, and 60 percent finer by weight, respectively. The $D_{50}$ and the coefficient of uniformity (Cu) were used to represent the characteristics, where Cu is defined as $\frac{D_{60}}{D_{10}}$.

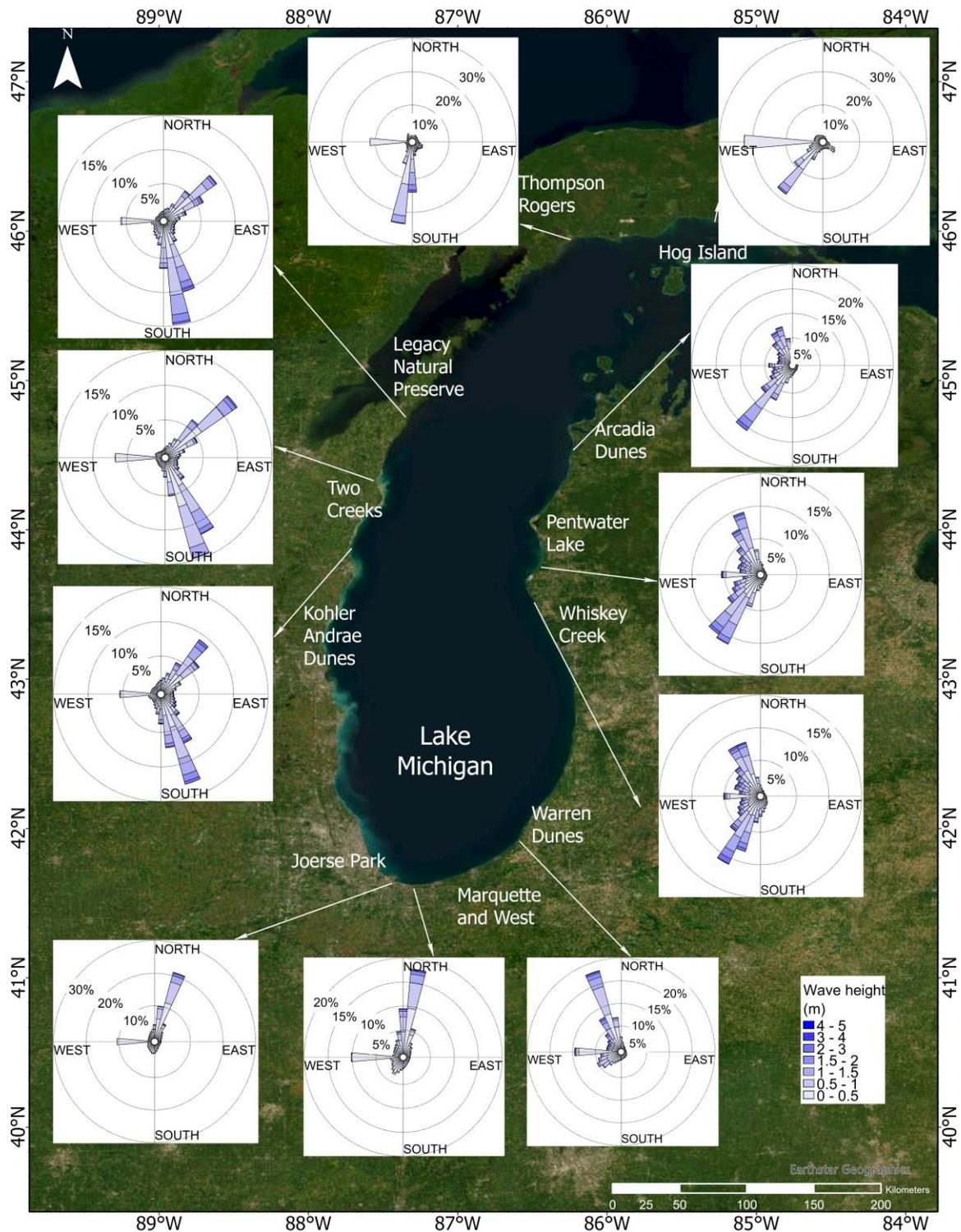

**Figure 4**. Wave roses for the study sites, for the study period (2013-2020). Roses indicate direction from which waves are arriving, with radial percentages indicating the fraction of waves from a given direction band. Colors within direction bands indicate the significant wave height distribution for the particular band.

2.5 Potential causative factors

Potential causative factors for shoreline changes, mentioned earlier in the introduction, were calculated for each site for the period between 2013 and 2020: wave power, storm wave energy, Wave bidirectionality index (WBI), longshore sediment transport divergence, number of days with ice cover, dimensionless fall velocity $\Omega$, Wave steepness divided by dimensionless fall velocity to the third power $\frac{S_o}{\Omega^3}$, the Bruun Rule slope, and the offshore slope. Detailed methodologies for calculating these factors are provided in the supplementary materials.

To assess the relationship between shoreline erosion and the factors, Pearson correlation analysis was performed. The correlation coefficient (r) was used to quantify the strength and direction of the relationship. A p-value threshold of 0.1 was applied to test the null hypothesis that no significant correlation exists between the variables. Correlations were considered statistically significant if the p-value was less than 0.1, indicating rejection of the null hypothesis.

## 3 Results

3.1 Satellite-derived shoreline changes

The analysis of shoreline changes revealed significant recession at all beaches in response to the water level increase from 2013 to 2020 (Figures 5, 6a; Table 2). The median shoreline recession across all sites ranged between 20 m and 62 m, with relative shoreline change magnitudes varying from 52% to 95% of the preexisting (2012) beach widths (Table 1).

Among all the sites, Whiskey Creek on the eastern shore of Lake Michigan experienced the most substantial shoreline recession of 62 m. Comparable magnitudes of shoreline recession were observed at Pentwater Lake (eastern Lake Michigan), Kohler Andrae (western Lake Michigan), Warren Dunes (southern Lake Michigan), Thompson Rogers and Hog Island (northern Lake Michigan). In contrast, Two Creeks and Legacy Natural Preserve on the western side of Lake Michigan, along with Jeorse Park Beach on the southern side of Lake Michigan, exhibited the smallest shoreline recession, of approximately 20 m. The shoreline recession for Marquette and West beaches (southern Lake Michigan) and Arcadia Dunes (eastern Lake Michigan) was close to the overall average for all sites, with a magnitude of around 35 m. Interestingly, the magnitudes of shoreline recession did not show a significant geographic pattern (Figures 5, 7).

**Table 2**. Magnitudes of shoreline changes, shoreline change standard deviation, coefficient of variance of shoreline changes (CV), beach face slope, and estimated shoreline erosion for all study sites. Coefficient of variance of shoreline changes is calculated as the shoreline standard deviation along each beach divided by the mean shoreline change.

| | Shoreline changes 2013 – 2020 (m) | Shoreline changes standard deviation along the beach (m) | Coefficient of variance of shoreline changes (CV) | Beach face slope (2012) (m/m) | Shoreline erosion 2013 – 2020 (m) |
|---|---|---|---|---|---|
| Whiskey Creek | 62±2.3 | 10.9 | 0.18 | 1/12.7±0.002 | 44.3 |

| | | | | |
|---|---|---|---|---|---|
| Pentwater Lake | 55±2.3 | 8.4 | 0.15 | 1/11.3±0.003 | 39.3 |
| Arcadia Dunes | 31±2.3 | 5.6 | 0.18 | 1/11.9±0.003 | 14.4 |
| Kohler Andrae Dunes | 57±2.3 | 5.2 | 0.09 | 1/9.6±0.002 | 43.8 |
| Two Creeks | 22±2.8 | 8.8 | 0.4 | 1/6.7±0.004 | 12.5 |
| Legacy Nature Preserve | 20±4.0 | 8.9 | 0.45 | 1/14.3±0.001 | 0.5 |
| Thompson Rogers | 54±4.0 | 8.6 | 0.16 | 1/25.3±0.001 | 15.4 |
| Hog Island | 48±4.0 | 10.5 | 0.22 | 1/26.0±0.001 | 17.8 |
| Warren Dunes | 48±2.3 | 15.2 | 0.32 | 1/18.3±0.001 | 22.7 |
| Marquette and West | 38±2.3 | 5.6 | 0.15 | 1/18.6±0.001 | 12.4 |
| Jeorse Park | 23.1±2.8 | 8.9 | 0.39 | 1/13.8±0.002 | 4.1 |

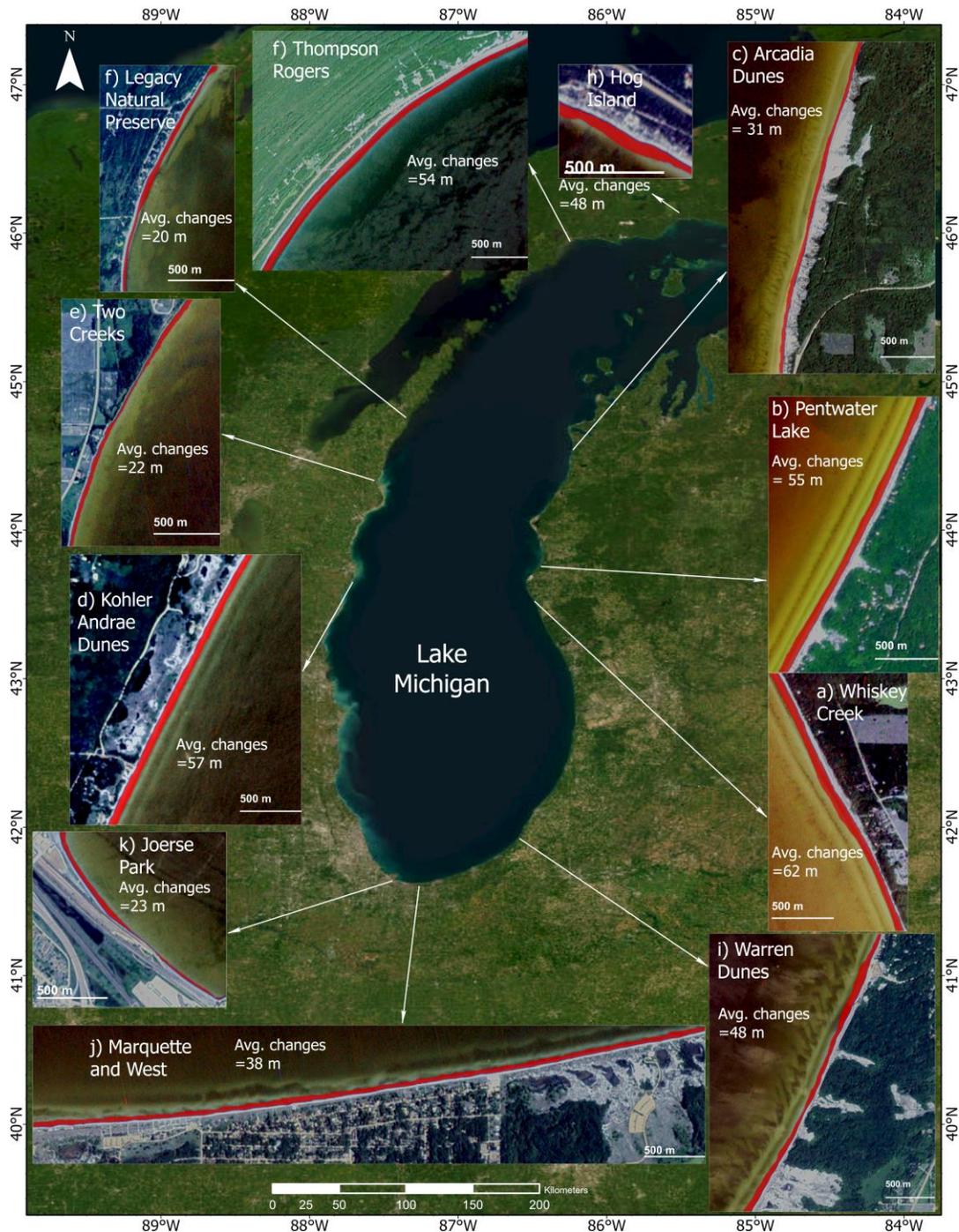

**Figure 5**. The detected shoreline changes for all the study sites. The red color indicates the eroded areas and its thickness is proportional to the erosion magnitude. Site images shown are 2020 Planetscope images.

In addition to median shoreline changes, substantial variability in shoreline changes was observed at some beaches (Table 2). Notably, Jeorse Park Beach, Legacy Natural Preserve, and Two Creeks exhibited considerable shoreline variability along the beach, with a coefficient of variance (CV) exceeding 0.4 (Table 2). This variability is clearly evident in the shoreline change visualization depicted in Figure 5. For instance, at Jeorse Park Beach, the northwestern section experienced an average shoreline recession of 30 m, while the eastern part saw a lesser recession of 12 m. All other sites had a CV of approximately 0.2, indicating more uniform shoreline changes within the study site (Table 2).

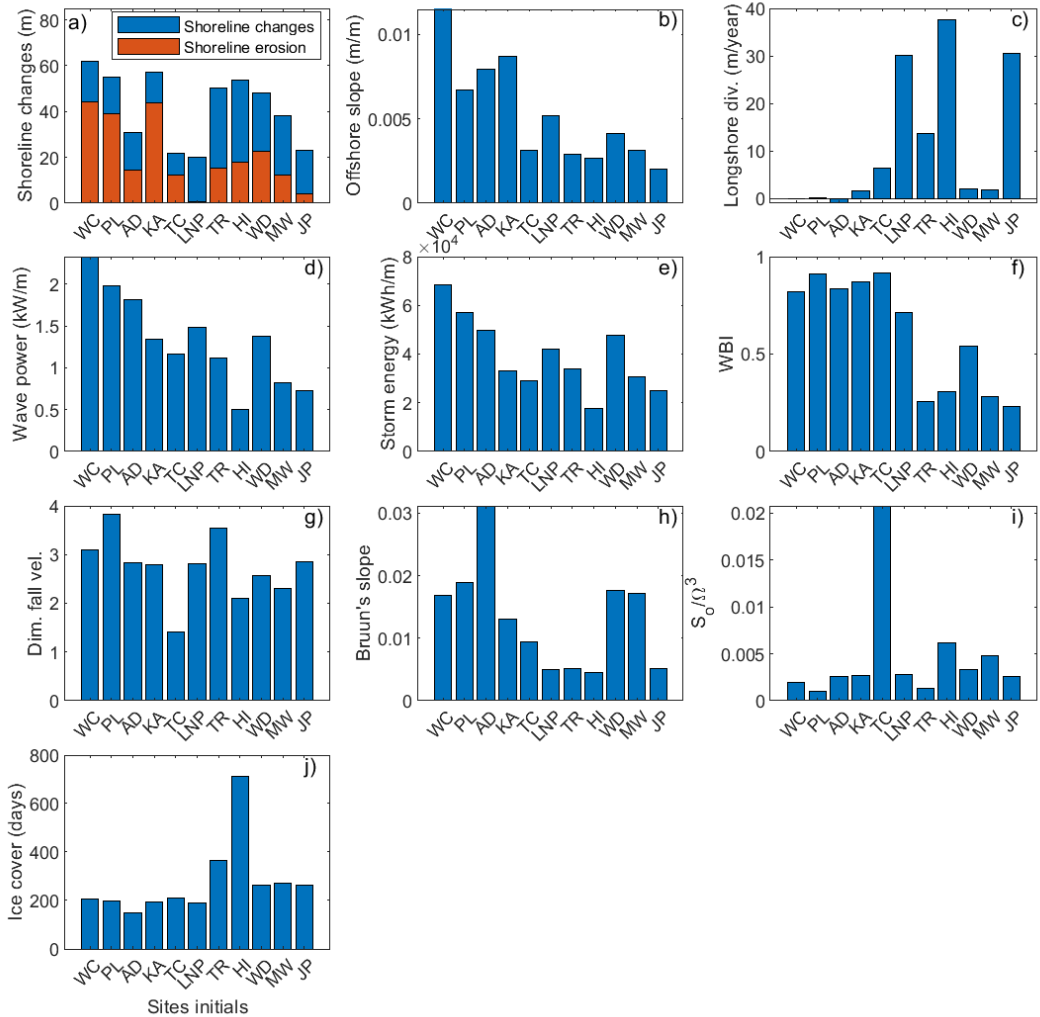

**Figure 6**. Calculated shoreline changes, shoreline erosion, and the potential causative factors calculated for all sites.

Similar to the shoreline changes, the average shoreline erosion exhibited significant variability across sites, ranging from minimal alterations to substantial shifts exceeding 40 m (Figures 6a, 7). Notably, Whiskey Creek and Pentwater Lake on eastern Lake Michigan, as well as Kohler Andrae on western Lake Michigan, experienced the most significant shoreline erosion from 2013 to 2020. Conversely, Legacy Natural Preserve on western Lake Michigan and Jeorse Park on southern Lake Michigan did not demonstrate significant shoreline erosion. For the other study sites, shoreline erosion ranged between 10 m and 25 m. When expressed as a percentage of the total change, all sites experienced a range of 25% to 99% of shoreline change due to passive inundation. The ratio for Whiskey Creek and Pentwater Lake on eastern Lake Michigan was 28%, which is close to the 20% reported by Hands (1980a) for the water level increase between 1967 and 1976 at similar sites on eastern Lake Michigan.

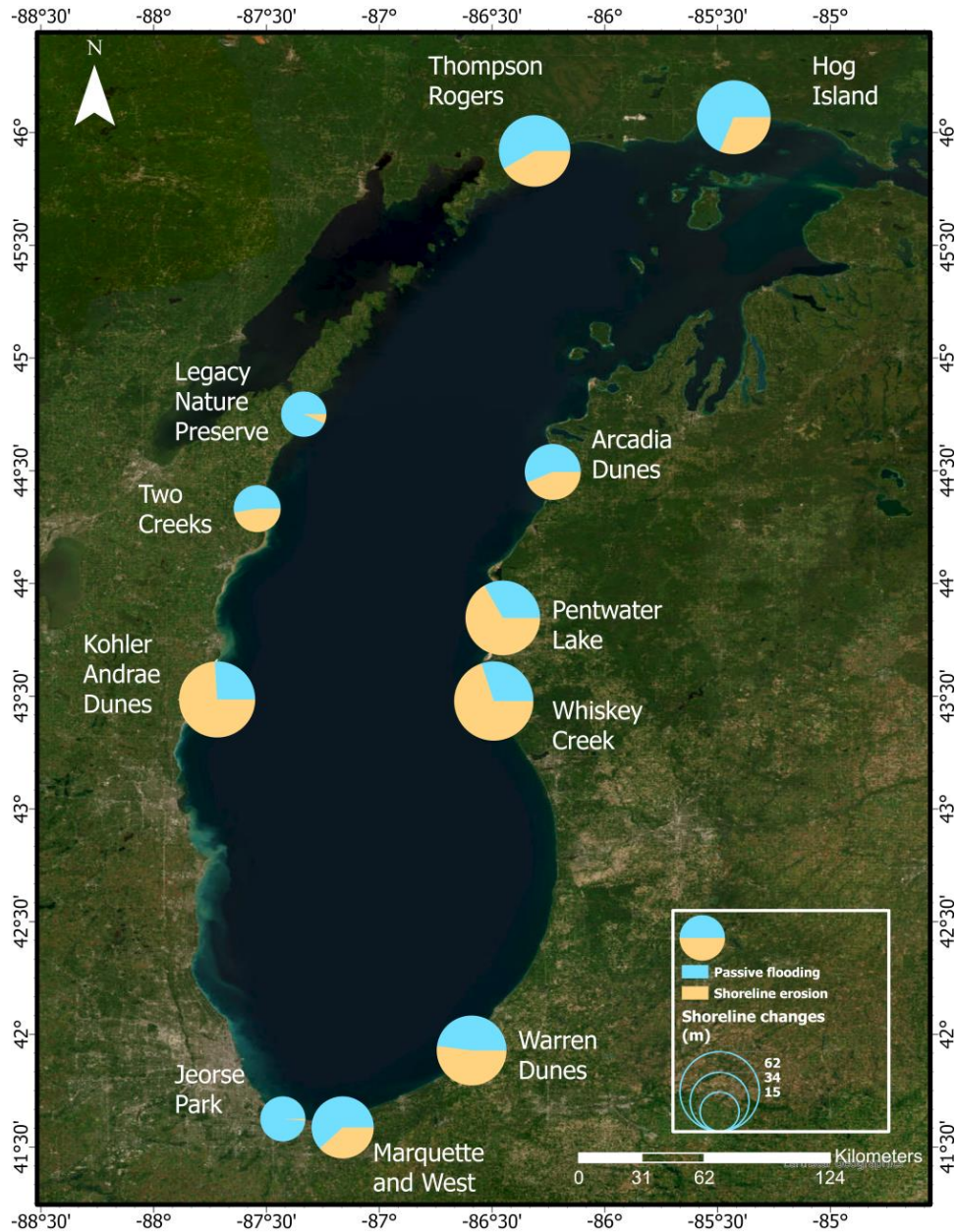

Figure 7 Shoreline changes measured at each site from satellite imagery. Shown are shoreline recessions, as well as the relative amounts attributable to passive flooding and shoreline erosion.

### 3.2 Beaches characteristics

Site visits and the analysis of beach characteristics revealed distinct features among the chosen sites that can influence their shoreline change behavior. Most sites exhibited fine to medium-grained sand with sizes ($D_{50}$), ranging between 0.23 to 0.35 mm, except for Two Creeks, which had coarse-grained sand with a $D_{50}$ of 0.7 mm (Table 1). Consequently, most sites had comparable dimensionless fall velocity ($\Omega$) and wave steepness divided by dimensionless fall velocity to the third power ($\frac{S_o}{\Omega^3}$), except for Two Creeks due to its coarse sediment size (Figures 6g, h).

The backshore conditions varied slightly between sites. Most sites featured vegetated sandy dunes of varying heights, except for Legacy Natural Preserve, which lacked clear dunes, and Two Creeks, which had an abrupt change from coarse sandy beach to clayey dunes (Table 1 and Figure 1). Beach width exhibited considerable variation among sites, with Whiskey Creek and Pentwater Lake having the widest beaches. Among the western sites, Two Creeks and Legacy Natural Preserve had the narrowest beaches, while Arcadia Dunes had the narrowest beach width and tallest dune height among the eastern sites. The beach width across all sites ranged from 25 to 87 m, with a mean width of 57 m (Table 1).

Beach face slopes varied significantly between sites, ranging from 1/27 to 1/8.6 (Table 2). Generally, the northern and southern beaches had more gentle slopes compared to the eastern and western beaches. Offshore slopes ranged between 1/500 and 1/86, with Whiskey Creek and Kohler Andrea having the steepest offshore slopes (Figure 6b), while the northern and southern beaches displayed more gentle ones. The Bruun Rule slopes were notably lower for Legacy Natural Preserve, Thompson Rogers, Hog Island, and Jeorse Park, while all other beaches had comparable Bruun Rule slopes, except for Arcadia Dunes due to its large berm height (Figure 6h).

Eastern Lake Michigan sites experienced the largest wave power and storm energy from 2013 to 2020, while the northern sites had the lowest values (Figures 3, 4; Figures 6d, e; Table S2). Moreover, wave roses in Figure 4 revealed that the northern and southern sites were mainly exposed to waves from a single dominant direction, whereas the eastern and western sites were subject to waves from two dominant directions, due to the north-south orientation of the lake influencing the wave energy distribution along its shores. This resulted in WBI values close to one at the east and west sites and near zero for the northern and southern sites (Figure 6f).

All sites, except Arcadia Dunes, displayed a sediment surplus with respect to longshore sediment transport divergence, indicating a potential net accretionary effect resulting from imbalances in longshore transport (Figure 6c). As defined, a higher positive longshore sediment transport divergence corresponds to a more pronounced accretion effect. Generally, the embayed beaches exhibited much larger positive longshore sediment transport divergence compared to the exposed beaches. Exposed beaches had an average divergence of $+1.2$ m$^3$/m$^2$/year, while embayed beaches had $+23.6$ m$^3$/m$^2$/year. Arcadia Dunes was the sole beach with a negative longshore sediment transport divergence of $-1.0$ m$^3$/m$^2$/year, countering the net accretionary trend observed at other sites.

Ice cover characteristics exhibited relatively consistent patterns across the sites, with limited spatial variation. The total number of ice cover days ranged between 150 and 273 days over the 7-year period from 2013 to 2020, except for the northern sites (Figure 6j). Thompson Rogers and Hog Island, the northern sites, experienced significantly longer periods of ice cover, with 365 and 714 days, respectively, during the same period.

### 3.3 Correlation analysis

Significant correlations were found between shoreline erosion and offshore slopes (r = 0.74), longshore sediment transport divergence (r = 0.59), wave power (r = 0.58), and storm energy (r = 0.55). Other factors, such as WBI, $\Omega$, the number of days with ice cover, $\frac{S_o}{\Omega^3}$, and the Bruun rule slope, were found to be statistically non-significant (p-value > 0.1).

The strong correlations between shoreline erosion and offshore beach slope, storm energy, longshore sediment transport divergence, and offshore wave power suggest that wave climate heterogeneity is a primary driver for the spatial variability in shoreline erosion (Figures 8 and 9). Sites with more gentle offshore slopes benefit from attenuated wave energy as waves approach the shore, as well as wave breaking and dissipation across a wider cross-shore area compared to steeper slopes (Van der Meer, 1990). In contrast, steeper offshore slopes allow a greater influx of wave energy into the surf zone, potentially mobilizing more sediment.

Spatially, eastern Lake Michigan experienced the highest storm energy and wave power from 2013 to 2020 (Figures 6d, e). This translated into greater shoreline changes at Whiskey Creek and Pentwater Lake but not at Arcadia Dunes. Hog Island, the northernmost site, was exposed to the least storm energy and wave power during this period, primarily due to prevailing ice cover during most of the winters, which mitigated the impact of winter storms. Nonetheless, Hog Island still experienced significant estimated erosion, averaging 17 m.

Longshore sediment transport divergence was found to have a stronger correlation with shoreline erosion than $\Omega$ and $\frac{S_o}{\Omega^3}$, highlighting the relative importance of longshore sediment transport over cross-shore sediment transport at these sites. Longshore sediment transport divergence is typically higher at sites with significant curvature due to the disparity in $\theta_b$ from the beginning to the end of the site. This can be observed in the high variability of shoreline changes within each beach, indicated by the coefficient of variance (CV), for beaches with large longshore sediment transport divergence (Table 2).

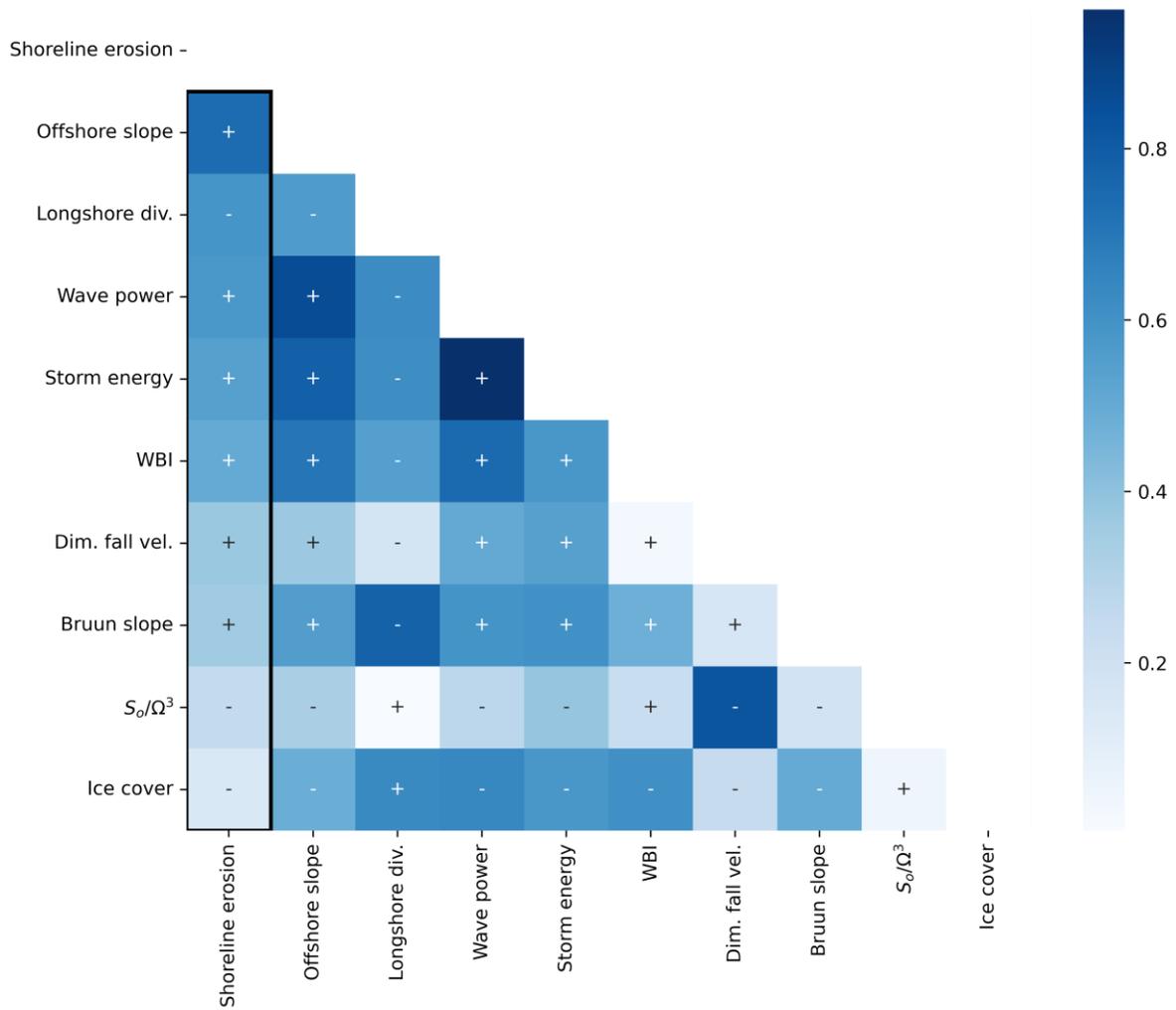

**Figure 8**. Pearson correlation coefficients between all the explanatory factors and the shoreline erosion (ranked from the highest to lowest with correct shoreline changes). Correlations between factors are also shown with positive and negative signs referring to positive and negative correlations.

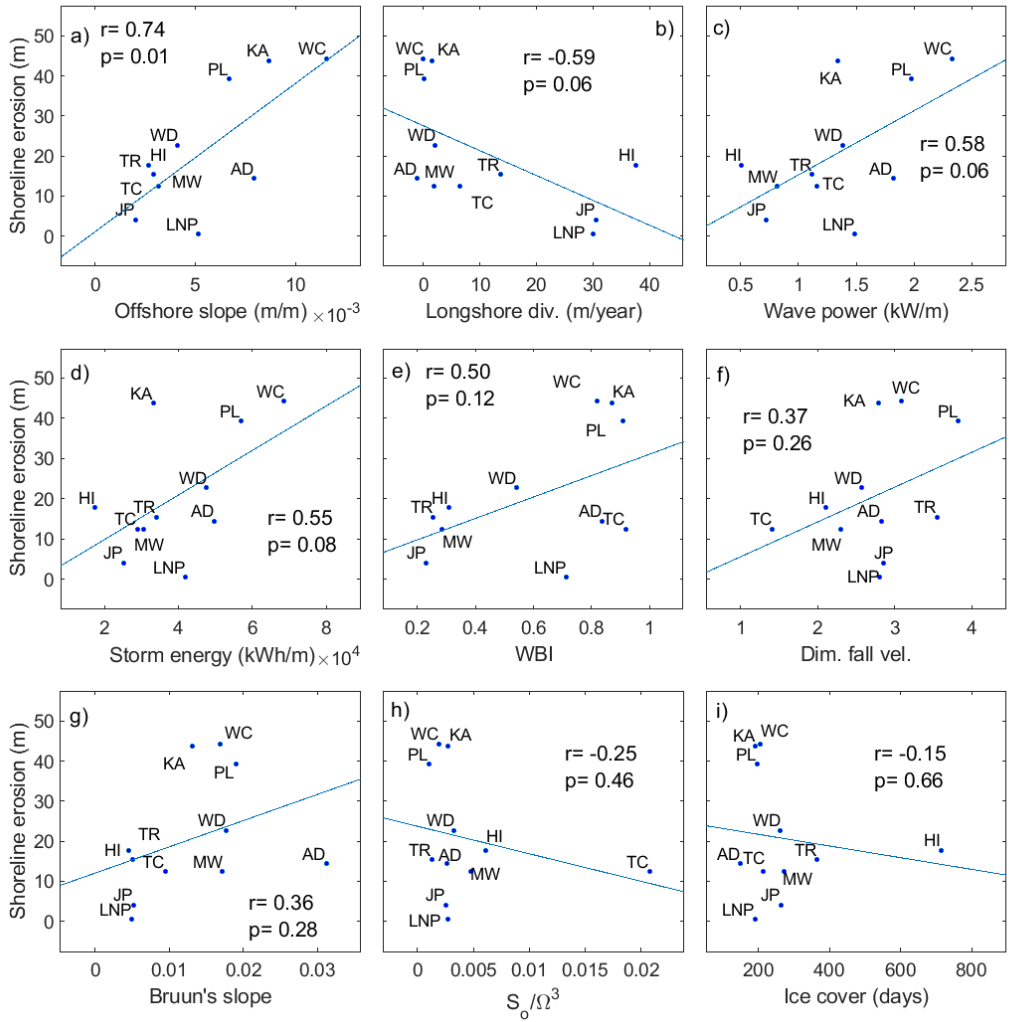

**Figure 9**. Scatter plots and calculated correlations between all potential causative factors and shoreline erosion.

## 4 Discussion

The results reveal substantial changes and variability in the shorelines of Lake Michigan in response to the water level increase between 2013 and 2020. After subtracting the passive flooding from total shoreline changes, the shoreline erosion still exhibited substantial spatial variability, indicating that other factors contributed to the observed differences. The correlation analysis identified a strong relationship between shoreline erosion and factors related to wave climate suggesting that the spatial variability in these factors is the primary reason for the differences in shoreline erosion between sites.

The findings underscore the importance of distinguishing between inundation and shoreline erosion when assessing the impact of water level increases on coastal areas with significant water level fluctuation. While all beaches experienced shoreline recession in response to the water level rise, some sites like Jeorse Park Beach and Legacy Natural Preserve showed minimal shoreline erosion, with most changes attributable to passive flooding. In contrast, beaches such as Whiskey Creek and Kohler Andrea experienced substantial erosion for the same water level increase. Sites primarily affected by passive flooding are likely to revert to their original positions once water levels decrease, whereas sites with significant shoreline erosion will require further attention. This highlights the importance of estimating shoreline erosion when designing coastal restoration plans.

Longshore sediment transport divergence was found to have a significant negative correlation with shoreline erosion, in contrast to traditional models like the Bruun Rule, which assume cross-shore profile adjustment to rising water levels but showed no significant correlation. This result challenges conventional expectations of how beaches respond to water level increases, suggesting that alongshore sediment dynamics play a more influential role. Positive longshore sediment transport divergence, indicating an accretional tendency, was observed at most sites, helping to offset erosion. The only exception was Arcadia Dunes, an exposed beach on the eastern shore of Lake Michigan with a sediment deficit. Despite these conditions, Arcadia Dunes experienced the least shoreline erosion (12 m) among all exposed sites, likely due to the continuous sand supply from the large (~80 m tall) dunes behind the beach, which mitigates shoreline recession. Additionally, the narrow beach width at Arcadia Dunes means that erosion almost reached the foredunes, which recedes at slower rates than shoreline changes due to the increase in water level (Davidson-Arnott & Bauer, 2021; van Dijk, 2021).

Although the ice cover factor was weakly correlated with shoreline erosion, it was strongly correlated with wave-related factors such as average wave power and storm wave energy. For instance, Hog Island in northern Lake Michigan experienced the lowest storm energy between 2013 and 2020 and the highest number of days with ice cover. Since ice cover occurs concurrently with many winter storms, it is believed to offer effective natural shoreline protection for coastal areas in the Great Lakes (BaMasoud & Byrne, 2012). However, with more than 70% of the Great Lakes' ice cover lost in recent decades (Wang et al., 2012), coastal areas are becoming increasingly vulnerable to storm activity, leading to heightened erosion risks. It is important to acknowledge that the ice cover examined in this study was not sufficiently detailed to capture the potential armoring effect of the ice on the beach face (BaMasoud & Byrne, 2012; Dodge et al., 2022). Despite the protective effect of ice cover against wave action, Hog Island still faced significant shoreline erosion (17 m), raising questions about the erosion mechanisms at beaches with similar conditions. One possible explanation may be the lack of sandy beaches and dunes around Hog Island, which could limit sediment availability.

The $\Omega$, $\frac{S_o}{\Omega^3}$, and WBI factors were found to have non-significant correlations with the shoreline erosion (Figures 8, 9-h, i, j). These factors have been associated with beach typology, erosion, shoreline changes, and rotation on ocean beaches (Davidson et al., 2013; Reeve et al., 2016; Wiggins et al., 2019). However, this was not the case for our study sites. The lack of correlation between $\Omega$ and $\frac{S_o}{\Omega^3}$, and shoreline erosion for the beaches studied here may be due to the narrower range of these parameters in the present study compared to ocean coasts, making it more challenging to identify statistical dependencies. Ocean coastlines typically experience a mix of long-crested swells and short wind waves, whereas Lake Michigan is primarily influenced by smaller, short-crested wind waves and only occasionally by longer storm waves. This results in a limited range of $\Omega$ in the Great Lakes compared to the broader range observed in oceanic settings, thereby diminishing the effectiveness of these factors in explaining the spatial variability of erosion in Lake Michigan (Theuerkauf et al., 2019). Yet, the positive correlation for $\Omega$ and the negative correlation for $\frac{s_0}{\Omega^3}$ suggests that the general trend aligns with expectations. Additionally, for WBI, both eastern and western sites in this study exhibited similar bidirectionality but differing shoreline changes, suggesting that bidirectionality may not be a significant factor.

While the strongly correlated factors are predominantly wave-related (e.g., wave energy and longshore sediment transport divergence) or influence wave climate (e.g., offshore slopes), our results indicate that shoreline changes result from a combination of various factors, and no single factor alone can explain the variability in shoreline changes. For instance, Whiskey Creek, which experienced the most substantial shoreline changes, had a combination of the highest wave power and storm energy, the steepest offshore slopes, and negligible sediment transport divergence. This combination led to the most significant shoreline erosion among all sites. Conversely, Arcadia Dunes, which had a negative longshore sediment transport divergence, benefited from the consistent sand supply from the large dunes, significantly altering its shoreline changes pattern. This underscores the necessity of more comprehensive site-specific assessments to further elucidate factors that regulate shoreline erosion.

The Bruun Rule slope, which represents the shoreline changes expected if the beach reached an equilibrium state in response to rising water levels, was found to have non-significant correlation with shoreline erosion. This was in contrast to the longshore sediment transport divergence which was found to be strongly correlated. Unlike the ocean, where sea level rise occurs over very long-time scales, potentially allowing for water level/beach equilibrium to be achieved, the Great Lakes experience rapid, large water level fluctuations, preventing equilibrium. The effective recession slopes, defined as the ratio of water level change to total shoreline recession (Hands, 1980; Troy et al., 2021), range from 1:14 to 1:45, falling between passive flooding slopes (beach face slopes) and Bruun Rule slopes (Table 3). This suggests that these beaches are undergoing non-equilibrium erosion. For example, the effective recession slope was closest to the beach face slope at sites with limited shoreline erosion (e.g., Jeorse Park and Legacy Natural Preserve) but aligned more closely with the Bruun Rule slope at sites with significant erosion (e.g., Whiskey Creek). This finding highlights the limitations of using the Bruun Rule to predict shoreline changes related to water level increases, particularly in the Great Lakes, but also along other shorelines where beach adjustment cannot keep pace with sea level rise.

Although current oceanic sea level rise rates are much lower than the water level increases observed in the Great Lakes, projections indicate that global sea levels could rise more than 2 meters by 2100 in some climate scenarios (Sweet et al., 2022), which is comparable to the rise seen in Lake Michigan-Huron from 2013 to 2020. This current study highlights the range of responses that should be expected from sea level rise along oceanic coasts, from passive flooding to Bruun's Rule recession (Table 3), especially for tideless and fetch-limited environments. This emphasizes the importance of including factors beyond sea level rise magnitude—similar to those considered in this study—to achieve more realistic predictions of future oceanic shoreline responses to sea level rise. However, a 2-meter sea-level rise over a century differs from a sub-decadal rise due to the extended period available for coastal processes to approach an equilibrium beach profile, as suggested by the Bruun rule. Additionally, the hydrodynamic differences between oceanic and lacustrine environments, such as wave energy, tidal influences, and storm surges, must be taken into account to fully understand and predict the coastal impacts of sea level rise in oceanic contexts.

**Table 3**. Different recession slopes for all study sites.

| | Beach face slope: passive flooding slope | Bruun's Rule slope | Effective recession slope |
|---|---|---|---|
| Whiskey Creek | 1/12.7 | 1:59 | 1:44.8 |
| Pentwater Lake | 1/11.3 | 1:53 | 1:39.7 |
| Arcadia Dunes | 1/11.9 | 1:32 | 1:22.3 |
| Kohler Andrae Dunes | 1/9.6 | 1:76 | 1:41.3 |
| Two Creeks | 1/6.7 | 1:105 | 1:15.7 |
| Legacy Nature Preserve | 1/14.3 | 1:202 | 1:14.6 |
| Thompson Rogers | 1/25.3 | 1:197 | 1:36.5 |
| Hog Island | 1/26.0 | 1:220 | 1:38.9 |
| Warren Dunes | 1/18.3 | 1:56 | 1:34.8 |
| Marquette and west | 1/18.6 | 1:58 | 1:27.6 |
| Jeorse Park | 1/13.8 | 1:193 | 1:14.1 |

It should be noted that the definition of shoreline position and shoreline erosion in this study differs from traditional geomorphological definitions. Typically, shoreline position is defined using markers such as the foredune toe, the edge of vegetation, and the boundary between the foreshore and backshore. In contrast, our definition of shoreline position as the transition from water to land inherently results in fluctuations due to wave activity and water level changes over short timescales, which contributes to increased uncertainty. However, fluctuations resulting in changes smaller than the pixel size (3-5 meters) are absorbed by the image resolution. Larger changes are further mitigated by averaging shoreline positions derived from multiple images taken in June and July of 2013 and 2020. This approach reduces the overall uncertainty in the calculated shoreline changes.

Similarly, traditional quantification of shoreline erosion often involves comparing topobathymetric surveys conducted before and after water level increases, but such surveys are typically unavailable, particularly for rural coastal areas. Our method is based on the hypothesis that passive flooding and shoreline erosion occur on different timescales: passive flooding happens almost instantaneously with water level rises, whereas morphological changes (erosion) take more time to manifest. Using initial profiles from the 2012 topobathymetric survey, we calculated the shoreline changes that would occur if no erosion had taken place (passive flooding). We then calculated actual shoreline changes from satellite images, attributing any changes exceeding those expected from passive flooding to morphological changes (erosion). This method serves as a proxy for assessing shoreline erosion on a large scale, but may differ from traditional methods in areas with complex beach topography. Therefore, traditional methods should still be employed for detailed site-specific assessments.

## 5 Conclusions

In this paper, shoreline responses at eleven Lake Michigan beaches after a record-setting water level rise between 2013 and 2020 were quantified and characterized, using high-resolution multispectral satellite images. A shoreline detection model was applied to the satellite images to accurately detect the shoreline position and calculate the shoreline changes. The analysis of shoreline changes revealed significant variations across the beaches, despite all being impacted by the same water level increase. To assess shoreline erosion at these sites, the passive inundation was subtracted from the total shoreline changes detected from the satellite images. Additionally, nine morphological and hydrodynamic parameters were investigated to discern the primary factors influencing the spatial variability of the shoreline erosion across the selected sites.

The results highlighted substantial variability in total shoreline changes, with all beaches experiencing recession ranging from 20 m to 62 m. Shoreline erosion varied considerably from site to site, with some sites experiencing erosion up to 44 m, while others showed minimal changes on average. Notably, Eastern Lake Michigan sites were most affected by the highest wave power and storm activity from 2013 to 2020. Beaches with steeper offshore slopes and higher wave energy demonstrated more pronounced shoreline erosion compared to those with milder slopes and lower energy. These findings underscore the crucial role of wave climate and beach characteristics in shaping shoreline responses to water level increases.

This study emphasizes the profound impact of the recent water level increase on Lake Michigan shorelines and highlights how diverse morphological and hydrodynamic conditions can influence shoreline response, even under the same water level rise. The results underscore the importance of distinguishing between immediately reversible changes (inundation) and changes that could be reversible over longer timescales, if at all (erosion) when assessing the impact of rising water levels on coastal regions. This distinction is essential for developing effective coastal restoration and management strategies to protect Great Lakes coastal areas from future water level fluctuations. Furthermore, the identified explanatory parameters can be used to pinpoint the most vulnerable beaches along the Great Lakes, guiding targeted preservation and protection efforts using inexpensive and readily available data. Finally, the study underscores the need for more detailed, site-specific research to better understand coastal erosion in response to the combined effects of water levels, ice cover, and local wave conditions. Such understanding is crucial for safeguarding the unique and valuable ecosystems and infrastructure along Great Lakes shorelines from future water level increase events.

### Acknowledgments

This work was supported in part by the Illinois-Indiana Sea Grant College Program, grant number NA18OAR4170082, and the Indiana Department of Natural Resources Lake Michigan Coastal Program, grant number NA20NOS4190036. HA also acknowledges support from the Lyles School of Civil Engineering. We acknowledge the use of imagery from the Smallsat Data Explore application (https://csdap.earthdata.nasa.gov), part of the NASA Commercial Smallsat Data Acquisition Program.

## Data statement

Satellite images used in this work can be downloaded from (https://www.planet.com/explorer/). Planetscope Scene and RapidEye Ortho Tile datasets were used in this study (Planet Labs, 2020). We got access to the satellite images from the Smallsat Data Explore application (https://csdap.earthdata.nasa.gov/signup/?_ga=2.86965151.641478145.1643651155-43285118.1641411107), part of the NASA Commercial Smallsat Data Acquisition Program (NASA, 2021).

Topo-bathymetric LiDAR data was extracted from (https://coast.noaa.gov/dataviewer/#/lidar/search/). The dataset name is (2012 USACE NCMP Topobathy Lidar DEM: Lake Michigan), and it is publicly available (NOAA, 2012).

Climate Forecast System Version 2 (CFSv2)5 dataset is available through (https://www.ncei.noaa.gov/access/metadata/landing-page/bin/iso?id=gov.noaa.ncdc:C00877), and it is publicly available (NOAA, 2011).

Water level data was downloaded from (https://tidesandcurrents.noaa.gov/), and it is publicly available. Calumet Harbor, IL (9087044) dataset was used in this study (NOAA, 2021).

Ice cover data was obtained from (https://www.glerl.noaa.gov/data/ice/) and it is publicly available (NOAA, 2022).

## References

Abdelhady, H. U., & Troy, C. D. (2023a). A reduced-complexity shoreline model for coastal areas with large water level fluctuations. *Coastal Engineering*, *179*, 104249. https://doi.org/10.1016/j.coastaleng.2022.104249

Abdelhady, H. U., & Troy, C. D. (2023b). Modeling Lake Michigan shoreline changes in response to rapid water level fluctuations. *Coastal Sediments*, 1332–1346. https://doi.org/10.1142/9789811275135_0123

Abdelhady, H. U., Troy, C. D., Habib, A., & Manish, R. (2022). A simple, fully automated shoreline detection algorithm for high-resolution multi-spectral imagery. *Remote Sensing*, *14*(3), 557. https://doi.org/10.3390/rs14030557

Assel, R. A., Wang, J., Clites, A. H., Bai, X., & Blank, R. (2013). Analysis of Great Lakes ice cover climatology: Winters 2006-2011. *NOAA Technical Memorandum GLERL-157 ANALYSIS OF GREAT LAKES ICE COVER CLIMATOLOGY*. https://repository.library.noaa.gov/view/noaa/11136/noaa_11136_DS1.pdf

Assel, R., Cronk, K., & Norton, D. (2003). Recent trends in Laurentian Great Lakes ice cover. *Climatic Change*, *57*(1–2), 185–204. https://doi.org/10.1023/A:1022140604052

BaMasoud, A., & Byrne, M. L. (2012). The impact of low ice cover on shoreline recession: A case study from Western Point Pelee, Canada. *Geomorphology*, *173–174*, 141–148. https://doi.org/10.1016/J.GEOMORPH.2012.06.004

Davidson, M. A., Splinter, K. D., & Turner, I. L. (2013). A simple equilibrium model for predicting shoreline change. *Coastal Engineering*, *73*, 191–202. https://doi.org/10.1016/j.coastaleng.2012.11.002

Davidson-Arnott, R. G. D., & Bauer, B. O. (2021). Controls on the geomorphic response of beach-dune systems to water level rise. *Journal of Great Lakes Research*, *47*(6), 1594–1612. https://doi.org/10.1016/J.JGLR.2021.05.006

Dean, R. G., & Dalrymple, R. A. (2001a). Coastal Processes with Engineering Applications. *Coastal Processes with Engineering Applications*. https://doi.org/10.1017/CBO9780511754500

Dean, R. G., & Dalrymple, R. A. (2001b). Coastal Processes with Engineering Applications. In *Cambridge University Press*. Cambridge University Press. https://doi.org/10.1017/cbo9780511754500

Dean, R. G., & Houston, J. R. (2016). Determining shoreline response to sea level rise. *Coastal Engineering*, *114*, 1–8. https://doi.org/10.1016/j.coastaleng.2016.03.009

Dodge, S. E., Zoet, L. K., Rawling, J. E., Theuerkauf, E. J., & Hansen, D. D. (2022). Transport properties of fast ice within the nearshore. *Coastal Engineering*, *177*, 104176. https://doi.org/10.1016/J.COASTALENG.2022.104176

Douglas, B. C., & Crowell, M. (2000). Long-term shoreline position prediction and error propagation. *Journal of Coastal Research*, *16*(1), 145–152.

Elias, J. E., & Meyer, M. W. (2003). Comparisons of undeveloped and developed shorelands, northern Wisconsin, and recommendations for restoration. *Wetlands*, *23*(4), 800–816. https://doi.org/10.1672/0277-5212(2003)023[0800:COUADS]2.0.CO;2

Gronewold, A. D., & Rood, R. B. (2019). Recent water level changes across Earth's largest lake system and implications for future variability. *Journal of Great Lakes Research*, *45*(1), 1–3. https://doi.org/10.1016/J.JGLR.2018.10.012

Hands, E. B. (1980). Changes in rates of shore retreat : Lake Michigan, 1967-76 / by Edward B. Hands. In *U.S. Army, Corps of Engineers, Coastal Engineering Research Center*. U.S. Army, Corps of Engineers, Coastal Engineering Research Center. https://doi.org/10.5962/bhl.title.47160

Hands, E. B. (1984). The Great Lakes as a test model for profile response to sea level changes. *This Digital Resource Was Created from Scans of the Print Resource.* https://erdc-library.erdc.dren.mil/jspui/handle/11681/12950

Hanrahan, J. L., Kravtsov, S. V., & Roebber, P. J. (2009). Quasi-periodic decadal cycles in levels of lakes Michigan and Huron. *Journal of Great Lakes Research*, *35*(1), 30–35. https://doi.org/10.1016/j.jglr.2008.11.004

Huang, C., Zhu, L., Ma, G., Meadows, G. A., & Xue, P. (2021). Wave climate associated with changing water level and ice cover in Lake Michigan. *Frontiers in Marine Science*, *8*, 746916. https://doi.org/10.3389/fmars.2021.746916

Jose, F., Martinez-Frias, J., Kilibarda, Z., & Kilibarda, V. (2022). Foredune and Beach Dynamics on the Southern Shores of Lake Michigan during Recent High Water Levels. *Geosciences 2022, Vol. 12, Page 151*, *12*(4), 151. https://doi.org/10.3390/GEOSCIENCES12040151

Karunarathna, H., Pender, D., Ranasinghe, R., Short, A. D., & Reeve, D. E. (2014). The effects of storm clustering on beach profile variability. *Marine Geology*, *348*, 103–112. https://doi.org/10.1016/J.MARGEO.2013.12.007

Kraus, N. C., Larson, M., & Kriebel, D. L. (1991). Evaluation of beach erosion and accretion predictors. *Coastal Sediments*, 572–587.

Lent, L. (2004). *Economics of the Shoreline: An Annotated Bibliography for the National Shoreline Management Study*. https://usace.contentdm.oclc.org/digital/collection/p16021coll2/id/1355/

MacIver, B. N., & Hale, G. P. (1970). *ASTM and EM 1110-2-1906 standards: LABORATORY SOILS TESTING*. U. S. Army Engineer Waterways Experiment Station. https://www.publications.usace.army.mil/portals/76/publications/engineermanuals/em_1110 -2-1906.pdf

Mattheus, C. R., Theuerkauf, E. J., & Braun, K. N. (2022). Sedimentary in-filling of an urban Great Lakes waterfront embayment and implications for threshold-driven shoreline morphodynamics, Montrose Beach, SW Lake Michigan. *Journal of Great Lakes Research*. https://doi.org/10.1016/J.JGLR.2022.06.008

Meadows, G. A., Meadows, L. A., Wood, W. L., Hubertz, J. M., & M. Perlin. (1997). The relationship between Great Lakes water levels, wave energies, and shoreline damage. *Bulletin of the American Meteorological Society*. https://doi.org/https://doi.org/10.1175/1520-0477(1997)078%3C0675:TRBGLW%3E2.0.CO;2

Miller, J. K., & Dean, R. G. (2004). A simple new shoreline change model. *Coastal Engineering*, *51*(7), 531–556. https://doi.org/10.1016/j.coastaleng.2004.05.006

NASA. (2021). *Commercial Smallsat Data Acquisition (CSDA) Program | Earthdata*. https://www.earthdata.nasa.gov/esds/csda

Nicholls, R. J., Birkemeier, W. A., & Lee, G. hong. (1998). Evaluation of depth of closure using data from Duck, NC, USA. *Marine Geology*, *148*(3–4), 179–201. https://doi.org/10.1016/S0025-3227(98)00011-5

NOAA. (2011). *Climate Forecast System Version 2 (CFSv2) Operational Forecasts [Dataset]*. https://www.ncei.noaa.gov/access/metadata/landing-page/bin/iso?id=gov.noaa.ncdc:C00877 (Data Access 2022-05-26)

NOAA. (2012). *NOAA: Data Access Viewer - Digital Coast - LiDAR [Dataset]*. https://coast.noaa.gov/dataviewer/#/lidar/search/ (Data Accessed 2022-05-20)

NOAA. (2021). *Water Levels - NOAA Tides & Currents [Dataset]*. Water Levels - NOAA Tides & Currents. https://tidesandcurrents.noaa.gov/waterlevels.html?id=9087044&type=Tide+Data&name=C alumet Harbor&state=IL [Data Accessed 2022-1-10]

NOAA. (2022). *Ice Cover: NOAA Great Lakes Environmental Research Laboratory - Ann Arbor, MI, USA [Dataset]*. https://www.glerl.noaa.gov/data/ice/ (Data Accessed 2021-10-03)

OCM Partners. (2012). *USACE NCMP Topobathy Lidar DEM: Lake Michigan*. https://www.fisheries.noaa.gov/inport/item/64407

Planet Labs, P. (2020). *Planet application program interface: In space for life on Earth [Dataset]*. https://www.planet.com/explorer/ (Data Accessed 2022-6-15)

Porst, G., Brauns, M., Irvine, K., Solimini, A., Sandin, L., Pusch, M., & Miler, O. (2019). Effects of shoreline alteration and habitat heterogeneity on macroinvertebrate community composition across European lakes. *Ecological Indicators*, *98*, 285–296. https://doi.org/10.1016/J.ECOLIND.2018.10.062

Putnam, J. A., & Johson, J. W. (1949). The dissipation of wave energy by bottom friction. *Eos, Transactions American Geophysical Union*, *30*(1), 67–74. https://doi.org/10.1029/TR030I001P00067

Reeve, D., Chadwick, A., & Fleming, C. (2016). *Coastal engineering: processes, theory and design practice* (pp. 99–103). Crc Press. https://doi.org/10.1007/978-94-007-6644-0_138-1

Saha, S., Moorthi, S., Pan, H. L., Wu, X., Wang, J., Nadiga, S., Tripp, P., Kistler, R., Woollen, J., Behringer, D., Liu, H., Stokes, D., Grumbine, R., Gayno, G., Wang, J., Hou, Y. T., Chuang, H. Y., Juang, H. M. H., Sela, J., … Goldberg, M. (2010). The NCEP climate forecast system reanalysis. *Bulletin of the American Meteorological Society*, *91*(8), 1015–1057. https://doi.org/10.1175/2010BAMS3001.1

Saha, S., Moorthi, S., Wu, X., Wang, J., Nadiga, S., Tripp, P., Behringer, D., Hou, Y. T., Chuang, H. Y., Iredell, M., Ek, M., Meng, J., Yang, R., Mendez, M. P., Van Den Dool, H., Zhang, Q., Wang, W., Chen, M., & Becker, E. (2014). The NCEP climate forecast system version 2. *Journal of Climate*, *27*(6), 2185–2208. https://doi.org/10.1175/JCLI-D-12-00823.1

Sweet, W. V., Hamlington, B. D., Kopp, R. E., Weaver, C. P., Barnard, P. L., Bekaert, D., Brooks, W., Craghan, M., Dusek, G., Frederikse, T., Garner, G., Genz, A. S., Krasting, J. P., Larour, E., Marcy, D., Marra, J. J., Obeysekera, J., Osler, M., Pendleton, M., … Zuzak, C. (2022). *2022: Global and regional sea level rise scenarios for the United States*.

Theuerkauf, E. J., & Braun, K. N. (2021). Rapid water level rise drives unprecedented coastal habitat loss along the Great Lakes of North America. *Journal of Great Lakes Research*, *47*(4), 945–954. https://doi.org/10.1016/j.jglr.2021.05.004

Theuerkauf, E. J., Braun, K. N., Nelson, D. M., Kaplan, M., Vivirito, S., & Williams, J. D. (2019). Coastal geomorphic response to seasonal water-level rise in the Laurentian Great Lakes: An example from Illinois Beach State Park, USA. *Journal of Great Lakes Research*, *45*(6), 1055–1068.

Troy, C. D., Cheng, Y.-T., Lin, Y.-C., & Habib, A. (2021). Rapid lake Michigan shoreline changes revealed by UAV LiDAR surveys. *Coastal Engineering*, *170*, 104008. https://doi.org/10.1016/J.COASTALENG.2021.104008

Van der Meer, J. W. (1990). Extreme shallow water conditions - Design curves for uniform sloping beaches - Delf Hydraulics Report H198. In *H0198*. Deltares (WL). https://repository.tudelft.nl/islandora/object/uuid%3A6f6fd932-6a87-4a3f-b18a-0ab77250d3ba

van Dijk, D. (2021). Foredune dynamics at a Lake Michigan site during rising and high lake levels. *Journal of Great Lakes Research*, *47*(6), 1581–1593. https://doi.org/10.1016/J.JGLR.2021.10.012

Vitousek, S., & Barnard, P. L. (2015, July). *A nonlinear, implicit one-line model to predict long-term shoreline change*. https://doi.org/10.1142/9789814689977_0215

Vitousek, S., Barnard, P. L., Limber, P., Erikson, L., & Cole, B. (2017). A model integrating longshore and cross-shore processes for predicting long-term shoreline response to climate change. *Journal of Geophysical Research: Earth Surface*, *122*(4), 782–806. https://doi.org/10.1002/2016JF004065

Vos, K., Harley, M. D., Splinter, K. D., Simmons, J. A., & Turner, I. L. (2019). Sub-annual to multi-decadal shoreline variability from publicly available satellite imagery. *Coastal Engineering*, *150*, 160–174. https://doi.org/10.1016/J.COASTALENG.2019.04.004

Wang, J., Bai, X., Hu, H., Clites, A., Colton, M., & Lofgren, B. (2012). Temporal and spatial variability of Great Lakes ice cover, 1973–2010. *Journal of Climate*, *25*(4), 1318–1329. https://doi.org/10.1175/2011JCLI4066.1

Wiggins, M., Scott, T., Masselink, G., Russell, P., & Valiente, N. G. (2019). Regionally-coherent embayment rotation: Behavioural response to bi-directional waves and atmospheric forcing. *Journal of Marine Science and Engineering*, *7*(4), 116. https://doi.org/10.3390/jmse7040116

Yang, T. Y., Kessler, J., Mason, L., Chu, P. Y., & Wang, J. (2020). A consistent Great Lakes ice cover digital data set for winters 1973–2019. *Scientific Data 2020 7:1*, *7*(1), 1–12. https://doi.org/10.1038/s41597-020-00603-1

Yates, M. L., Guza, R. T., & O'Reilly, W. C. (2009). Equilibrium shoreline response: Observations and modeling. *Journal of Geophysical Research: Oceans*, *114*(9). https://doi.org/10.1029/2009JC005359